# Test of Nonextensive Statistical Mechanics by Solar Sound Speeds

Du Jiulin[*]

*Department of Physics, School of Science, Tianjin University, Tianjin 300072, China*

To check the validity of the theory of nonextensive statistical mechanics, we have investigated the nonextensive degree of the solar interior and have tried to find the experimental evidence by helioseismological measurements that $q$ is different from unity. We are able to derive a parameter for providing a lower limit to the nonextensive degree inside the sun that can be uniquely determined by the solar sound speeds measured by helioseismology. After calculating the parameter by using the solar sound speeds, we get the lower limit of $(1-q) \geq 0.1902$ for all solar radii between $0.15R_\odot$ and $0.95R_\odot$ and $(1-q) \approx 0.4$ for the out layers, $0.75R_\odot \leq r \leq 0.95R_\odot$. Thus, the result that the nonextensive parameter $q$ is significantly different from unity in the sun has received the support by the experiment measurements for the solar sound speeds in the helioseismology.

---

[*] Email Address: jiulindu@yahoo.com.cn

Nonextensive statistical mechanics (NSM) has been developed recently as a very useful tool to describe the complex systems whose properties cannot be exactly described by Boltzmann-Gibbs (B-G) statistical mechanics [1]. It has been applied extensively to deal with varieties of interesting problems in the field of astrophysics where the systems are known to need NSM due to the long-range nature of gravitational interactions, such as the properties of self-gravitating systems [2,3], Antonov problem [4], black hole [5], solar wind intermittency [6], stellar polytrope [7], Jeans criterion [8], non-Maxwell distribution in plasma [9], galaxy clusters [10], dark matter and dark energy [11] and etc.

NSM is based on the nonextensive entropy proposed by Tsallis as one generalization of B-G entropy. With a parameter $q$ different from unity, the nonextensive entropy is introduced [12] by

$$S_q = \frac{k}{1-q}\left(\sum_i p_i^q - 1\right) \tag{1}$$

where $k$ is the Boltzmann constant, and $p_i$ is the probability that the system under consideration is in its $i$-th configuration. A deviation of the parameter $q$ from unity is considered as describing the degree of nonextensivity of the system. In other words, (1-$q$) is a measure of the lack of extensivity of the system. The B-G entropy is recovered from $S_q$ only if we take $q$=1. For the canonical ensemble, Eq.(1) yields the four versions of power law distribution when considering four possible choices for the expectation values [13]. For instance, for the Tsallis-Mendes-Plastino one [14] we have

$$p_i \sim \left[1-(1-q)\beta(\varepsilon_i - U_q)\Big/\sum_i p_i^q\right]^{1/(1-q)} \tag{2}$$

and for the Curado-Tsallis one [15] we have

$$p_i \sim \left[1-(1-q)\beta\varepsilon_i\right]^{1/(1-q)} \tag{3}$$

where $\beta = 1/kT$ is the Lagrange parameter associated with the internal energy. These versions are now known to work equivalently [13].

Within the framework of NSM, many traditional theories in B-G statistical mechanics have been generalized for $q\neq 1$ and they have been applied to varieties of interesting fields [1]. To check the validity of the nonextensive theories, it has been urgently expected to be able to provide the evidence by the experiment fact that $q$ is different from unity. In this aspect, the nonextensive parameter was estimated at $|q-1| < 2.08 \times 10^{-5}$ from the primordial helium abundance in the early universe [16] and at $|q-1| < 3.6 \times 10^{-5}$ by the data from the cosmic microwave background radiation [17]. Up to now, however, we have not given any conclusive evidences by the fact that $q$ is significantly different from unity. In this letter, we try to find one of such evidences from the solar sound speeds measured by helioseismology.

As we know, a star is open system being in nonequilibrium state, which is usually in the hydrostatic equilibrium but not in thermal equilibrium, where the self-gravitating long-range interactions play an important role. Self-gravitating systems are generally thought to behavior nonextensively due to the long-range nature of gravitational forces. Strictly speaking, we should use the nonequilibrium and nonextensive statistical theory to study its thermodynamic properties. It is found that Tsallis equilibrium distribution can describe the hydrostatic equilibrium of the self-gravitating system such as stars [18]. When we apply B-G statistics to study the thermodynamic properties of the nonequlibrium system, we conventionally have to turn to the so-called "assumption of local equilibrium" for the statistical description of the system. This assumption requires the variation of a thermodynamic quantity in the scope of the mean free path of particles to be greatly less than the mean value of the same quantity within the same scope. Therefore, the standard Maxwell-Boltzmann (M-B) distribution is applied to local description for the gas inside a star. When we consider the non-local correlation in the nonequilibrium system, the M-B distribution needs to be generalized.

The generalized M-B distribution can be derived for $q \neq 1$ using a rigorous treatment based on the nonextensive formulation of Boltzmann equation and *H*-theorem [19], or through a simple nonextensive generalization of the Maxwell ansatz [20], or maximizing Tsallis entropy under the constraints imposed by normalization and the energy mean value [21]. When we consider a self-gravitating system with particles interacting via the gravitational potential $\varphi$, the generalized M-B distribution is given [3, 19, 20, 22, 27] by

$$f_q(\mathbf{r},\mathbf{v}) = nB_q \left(\frac{m}{2\pi kT}\right)^{\frac{3}{2}} \left[1-(1-q)\frac{m\mathbf{v}^2}{2kT}\right]^{\frac{1}{1-q}} \tag{4}$$

where $m$ is the mass of each particle, $T$ is the temperature, $n$ is the particle number density, $B_q$ is the $q$-dependent normalized constant. The standard M-B distribution is correctly recovered from Eq.(4) when we take $q=1$. Eq.(4) has been applied to investigate many interesting problems in astrophysics and the other fields [6, 22]. When discussing the true nature of the nonextensive parameter $q$, exactly the relation between $q$, the temperature gradient $\nabla T$ and the gravitational acceleration $\nabla\varphi$ is determined from Eq.(4) in combination with the generalized Boltzmann equation and *H*-theorem [3]. It is given by the following equation,

$$k\nabla T + (1-q)m\nabla\varphi = 0 \tag{5}$$

This equation provides an interesting physical interpretation for the nonextensive parameter $q \neq 1$: $q$ is not unity if and only if $\nabla T$ is not equal to zero, i.e., the deviation of $q$ from unity is closely related to the nonisothermal degree of the system. In other words, when a self-gravitating system is not at thermal equilibrium (almost all the self-gravitating systems at stable states are always so), the generalized M-B distribution Eq.(4) can be reasonably applied to describe the thermodynamic properties of the system *in the nonequilibrium stationary state*. So, to the extent Eq.(5) is at the base of *equilibrium of the nonextensive system* under the control of gravitational potential $\varphi$.

It is worth notice that Eq.(5) points to tests of NSM due to the measurablility of $\nabla T$ and $\nabla\varphi$ by experiments. It is strongly recommend using the nature data of stellar physics to estimate the value of $q$. We now apply Eq.(5) to the interior of a star. A star is generally regarded as the system of one dimension under spherical symmetry [23]. So, Eq.(5) is written as

$$-\frac{k}{\mu m_H}\frac{dT}{dr} = (1-q)\frac{d\varphi}{dr} = (1-q)\frac{GM(r)}{r^2} \tag{6}$$

or, to bring out explicitly the nonextensive degree parameter, equivalently,

$$1-q = -\frac{k}{\mu m_H}\frac{dT}{dr}\bigg/\frac{GM(r)}{r^2} \tag{7}$$

where $d\varphi/dr = GM(r)/r^2$ is the gravitational acceleration at the radius $r$, $G$ is the gravitational constant, $M(r)$ is the mass interior to a sphere of radius $r$. The mass $m$ is now defined by $\mu m_H$, $m_H$ is the mass of the hydrogen atom, and $\mu$ is the mean molecular weight that is defined by

$$\mu = [0.5 + 1.5X + 0.25Y]^{-1} \tag{8}$$

if the abundances of hydrogen and helium are represented by $X$ and $Y$, respectively.

The sun is perhaps among the best-understood stars today. In particular helioseismology has been applied to probing the interior structure and dynamics of the sun with ever-increasing precision, providing a well-calibrated laboratory in which physical processes and properties can be studied under conditions that are unattainable on Earth. For the solar interior, the sound speed can be introduced by

$$v_s = \sqrt{\Gamma_1 \frac{kT}{\mu m_H}} \tag{9}$$

where $\Gamma_1 \approx 5/3$ is the adiabatic index [24]. So, Eq.(7) is further written as

$$1-q = -\frac{k}{m_H}\left[\frac{d}{dr}\left(\frac{T}{\mu}\right) + \frac{T}{\mu^2}\frac{d\mu}{dr}\right]\bigg/\frac{GM(r)}{r^2}$$
$$= \left[-\frac{2v_s}{\Gamma_1}\frac{dv_s}{dr} - \frac{kT}{\mu^2 m_H}\frac{d\mu}{dr}\right]\bigg/\frac{GM(r)}{r^2} \tag{10}$$

As we well know, the mean molecular weight inside the sun decreases monotonously from the center outward and its gradients ($d\mu/dr$) are always non-positive everywhere for all solar radii. By eliminating the second term in Eq.(10), we obtain the inequality,

$$1-q \geq -\frac{2v_s}{\Gamma_1}\frac{dv_s}{dr}\bigg/\frac{GM(r)}{r^2} \tag{11}$$

where the equality sign is tenable only if $(d\mu/dr) = 0$. This inequality provides a lower limit to $(1-q)$ for the solar interior. Thus,

$$1-q \geq (1-q)^* \tag{12}$$

where the minimum nonextensive degree parameter, $(1-q)^*$, can be uniquely determined by the solar sound speeds measured in helioseismology,

$$(1-q)^* = -\frac{6}{5} v_s \frac{dv_s}{dr} \bigg/ \frac{GM(r)}{r^2} \tag{13}$$

In order to obtain the evidences by the experimental fact that $q$ is different from unity, we now come to the calculations for the values of $(1-q)^*$ by the solar sound speeds. It has been shown that the calculated sound speeds for the standard solar model (BP2000) and the helioseismologically measured sound speeds are the excellent agreement everywhere in the sun [25]: the rms fractional difference between the calculated and the measured sound speeds is 0.10% for all solar radii between $0.05R_\odot$ and $0.95R_\odot$ and is 0.08% for the deep interior region, $r \leq 0.25R_\odot$. For this reason we will use the BP2000 sound speeds as the measured ones in helioseismology, which are not any significant effects on our present results. The accurate data of the solar sound speeds for BP2000 are available by the full numerical details at the web site http://www.sns.ias.edu/~jnb .

Table 1 presents some numerical details of the solar interior of BP2000 and our calculated results. The first three columns provide the partial physical variables from BP2000: the radius (in units of $R_\odot$), the mass included in the interior to the sphere of the radius $r$ (in units of $M_\odot$), and the BP2000 sound speeds (in units of cm s$^{-1}$). The last three columns give the mean molecular weight $\mu$, the sound speed gradients $dv_s/dr$ (in units of s$^{-1}$) and the minimum nonextensive degree parameter, $(1-q)^*$, where $\mu$ is given using Eq.(8), $dv_s/dr$ is obtained using the detailed numerical description of the BP2000 sound speeds and $(1-q)^*$ is determined by using Eq.(13).

Fig.1 illustrates the distribution of values of the minimum nonextensive degree parameter, $(1-q)^*$, in the sun. The distribution of $(1-q)^*$ can be roughly divided into four regions for all solar radii. The first region is extending from the center to about $0.2R_\odot$ where $(1-q)^*$ increases gradually from a negative value at the center, it peaks with 0.2335 at $r = 0.20107R_\odot$, and then it has a little fall slowly. The second region is between $0.3R_\odot$ and $0.6R_\odot$ where $(1-q)^*$ waves approximately about 0.2. The third region is about from $0.6R_\odot$ to $0.71377R_\odot$ where $(1-q)^*$ increases further, it peaks again with 0.4330 at a radius of $r =0.71377R_\odot$ and then it has a small fall. The fourth one is about from $0.75R_\odot$ to $0.95R_\odot$ where $(1-q)^*$ keeps basically at the value of 0.4.

It is worth noticeable that $(1-q)^*$ is basically 0.4 in the region everywhere from $0.75R_\odot$ to $0.95R_\odot$, which we find to be exactly equal to the values of the nonextensive degree parameter, $(1-q)$, because the mean molecular weight $\mu$ is exactly uniform in this region and the second term on the right hand side of Eq.(10) is exactly zero. We therefore come to the conclusion that the nonextensive parameter $q$ is about 0.6 in the regions extending from $0.75R_\odot$ to $0.95R_\odot$, which is quite different from unity.

Thus, the result that the nonextensive parameter $q$ is significantly different from unity in the sun has received the support by the experiment measurements for the solar sound speeds in the helioseismology. In terms of the express for the nonextensive parameter $q$, Eq.(7), the nonextensivity represents the non-local spatial correlations within the gas with self-gravitating long-range interactions when being at the nonequilibrium stationary state. While the generalized M-B distribution, Eq.(4), can be reasonably applied to the non-local description for the thermodynamic properties of the long-range interacting system being in the nonequilibrium stationary state.

Summarizing, to check the validity of the theory of NSM, we have investigated the nonextensive degree of the solar interior and have tried to find the experimental evidence by the helioseismological measurements that $q$ is different from unity. We have been able to derive a parameter, $(1-q)^*$, for providing a lower limit to the nonextensive degree inside the sun that can be uniquely determined by the solar sound speeds measured in helioseismology. After calculating the parameter by using the solar sound speeds, we find the lower limit of $(1-q) \geq 0.1902$ for all solar radii between $0.15R_\odot$ and $0.95R_\odot$ and $(1-q) \approx 0.4$ for the out layers, $0.75R_\odot \leq r \leq 0.95R_\odot$. Although we use the BP2000 sound speeds as the measured ones in helioseismology for calculating the minimum nonextensive degree parameter, $(1-q)^*$, yet these will be not any significant effects on our present results because the rms fractional difference between the BP2000 calculated and the helioseismologically measured sound speeds is 0.10% for all solar radii between 0.05$R_\odot$ and 0.95$R_\odot$ and is 0.08% for the deep interior region, $r \leq 0.25R_\odot$, which shows the excellent agreement between the tow everywhere in the sun [25] (or visit the web site http://www.sns.ias.edu/~jnb ), proving the BP2000 sound speeds to be highly reliable.

**Note added**

In the paper [26], a test of M-B distribution with helioseismology was discussed by a modified statistics. Paper [27] dealt with a nonlinear kinetics on generalized Boltzmann equation, where Eq.(4) was derived as one of the results. Paper [28] also discussed a generalized Boltzmann equation.

I thank the project of "985" Program of TJU of China for the financial support.

Fig.1. Distribution of the minimum nonextensive degree parameter, $(1-q)^*$, determined by the sound speeds in the sun. These results are obtained from Table 1.

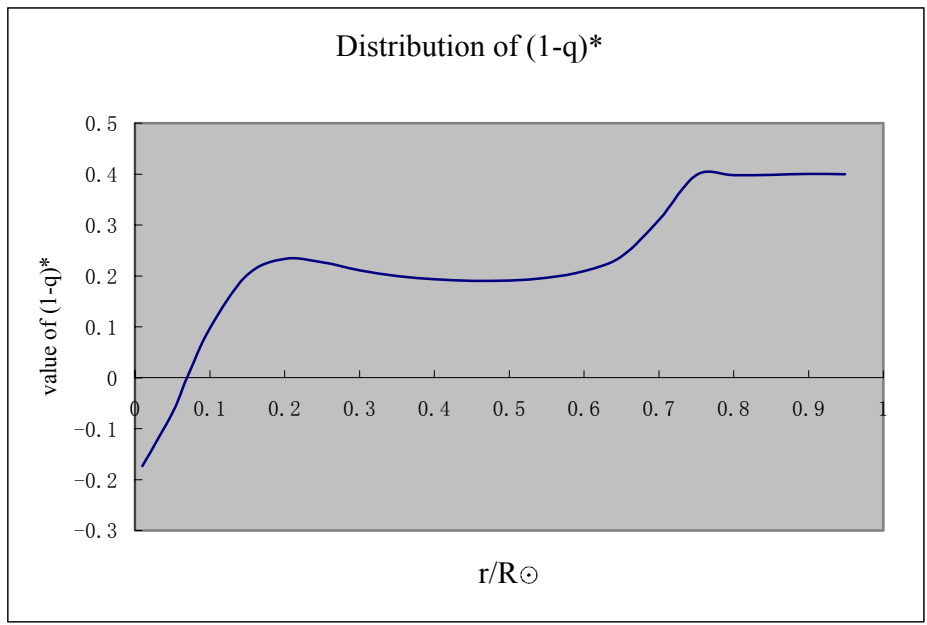

Table 1. Solar sound speed, sound speed gradient, mean molecular weight, and the minimum nonextensive degree parameter, $(1-q)^*$, as derived from the BP2000 solar sound speeds

| $r/R_\odot$ | $M/M_\odot$ | $v_s$ | $\mu$ | $dv_s/dr$ | $(1-q)^*$ |
|---|---|---|---|---|---|
| 0.01000 | 0.0001081 | 5.05656E+07 | 0.8532 | 8.4605E−05 | −0.1733 |
| 0.05034 | 0.0124165 | 5.10132E+07 | 0.7955 | 1.4879E−04 | −0.0678 |
| 0.06961 | 0.0301280 | 5.11219E+07 | 0.7578 | 0.0000E-00 | 0.0000 |
| 0.10006 | 0.0762478 | 5.07757E+07 | 0.7042 | −3.3330E−04 | 0.0973 |
| 0.15018 | 0.1925732 | 4.86931E+07 | 0.6494 | −8.0941E−04 | 0.2022 |
| 0.20107 | 0.3393826 | 4.54218E+07 | 0.6253 | −9.8517E−04 | 0.2335 |
| 0.25089 | 0.4841900 | 4.20378E+07 | 0.6171 | −9.4609E−04 | 0.2265 |
| 0.30016 | 0.6099028 | 3.89736E+07 | 0.6143 | −8.3665E−04 | 0.2110 |
| 0.35075 | 0.7132949 | 3.62308E+07 | 0.6124 | −7.2889E−04 | 0.1995 |
| 0.40067 | 0.7907148 | 3.38604E+07 | 0.6114 | −6.4093E−04 | 0.1932 |
| 0.45055 | 0.8480440 | 3.17676E+07 | 0.6106 | −5.7143E−04 | 0.1902 |
| 0.50063 | 0.8902432 | 2.98790E+07 | 0.6100 | −5.1831E−04 | 0.1910 |
| 0.55083 | 0.9211915 | 2.81372E+07 | 0.6094 | −4.8374E−04 | 0.1964 |
| 0.60081 | 0.9438189 | 2.64813E+07 | 0.6090 | −4.7327E−04 | 0.2099 |
| 0.64986 | 0.9602980 | 2.48312E+07 | 0.6084 | −4.9740E−04 | 0.2379 |
| 0.70042 | 0.9730081 | 2.29284E+07 | 0.6021 | −6.1235E−04 | 0.3101 |
| 0.71377 | 0.9758090 | 2.23201E+07 | 0.5989 | −8.4817E−04 | 0.4330 |
| 0.75033 | 0.9826089 | 2.03346E+07 | 0.5989 | −7.8089E−04 | 0.3985 |
| 0.80103 | 0.9900343 | 1.75530E+07 | 0.5989 | −7.9843E−04 | 0.3979 |
| 0.85023 | 0.9950969 | 1.47421E+07 | 0.5989 | −8.4980E−04 | 0.3986 |
| 0.90020 | 0.9982619 | 1.16042E+07 | 0.5989 | −9.7051E−04 | 0.4004 |
| 0.94896 | 0.9997083 | 7.87052E+06 | 0.5989 | −1.2871E−03 | 0.3997 |